\documentclass[a4paper]{article}

\usepackage{scicite}

\usepackage{times}

\usepackage{algorithm}
\usepackage{algorithmic}
\usepackage{graphicx}
\usepackage{multirow}
\usepackage{url}

\newenvironment{sciabstract}{%
\begin{quote} \bf}
{\end{quote}}

\newcounter{lastnote}

\title{Knowledge model: a method to evaluate an individual's knowledge quantitatively}

\author
{Gangli Liu$^{1\ast}$\\
	\\
	\normalsize{$^{1}$Department of Computer Science and Technology, Tsinghua University, Beijing, China}\\
	\\
	\normalsize{$^\ast$E-mail:  gl-liu13@mails.tsinghua.edu.cn.}
}

\date{}

\begin{document} 

% Double-space the manuscript.

\baselineskip24pt

% Make the title.

\maketitle

% Place your abstract within the special {sciabstract} environment.

\begin{sciabstract}
  As the quantity of human knowledge increasing rapidly, it is harder and harder to evaluate a knowledge worker's knowledge quantitatively. There are lots of demands for evaluating a knowledge worker's knowledge. For example, accurately finding out a researcher's research concentrations for the last three years; searching for common topics for two scientists with different academic backgrounds; helping a researcher discover his deficiencies on a research field etc. This paper proposes a method named knowledge model to evaluate a knowledge worker's knowledge quantitatively without taking an examination. It records and analyzes an individual's each learning experience, discovering all the involved knowledge points and calculating their shares by analyzing the text learning contents with topic model. It calculates a score for a knowledge point by accumulating the effects of one's all learning experiences about it. A preliminary knowledge evaluating system is developed to testify the practicability of knowledge model.
\end{sciabstract}

\section{Introduction}
The amount of human knowledge is rapidly increasing. Almost every discipline has been subdivided into lots of sub-disciplines. In information age, humans, especially knowledge workers, need to keep learning during their whole lives. There are a lot of demands for knowledge workers to estimate their knowledge quantitatively. The following are some examples:
\begin{itemize}
\item A computer engineer wants to estimate how much he has obtained the collection of concepts and algorithms of the curriculum ``Information Retrieval''; 
\item A researcher wants to predict how much he will understand the contents of a lecture just from its poster, a subsequent decision of whether to attend it will be made based on the prediction; 
\item Two researchers with different academic backgrounds want to find out the set of knowledge points on which they both have a solid understanding, these knowledge points can serve as the starting point of an academic communication. 
\item A scientist wants to have a quantitative evaluation of his research concentrations for the last three years.
\end{itemize}
Most of an individual's knowledge is obtained from postnatal learning. By recording and analyzing one's learning history, it is possible to estimate his knowledge quantitatively.

\subsection{Classification of an individual's activities}
To analyze one's learning history, an individual's daily activities are classified into two categories: learning activities and non-learning activities. Learning activities are those which are related to at least one piece of knowledge. The definitions of knowledge and non-knowledge will be explained in section 1.4.1.
Examples of learning activities are reading books, taking courses, discussing with someone about a piece of knowledge etc.

\subsection{Capturing the text learning contents}
Most learning processes can be associated with a piece of learning material. For example, reading a book, the book is the learning material. Taking a course or having a discussion, the course and discussion contents can be regarded as the learning material. Some of the learning materials are text or can be converted to text. For example, discussing about a piece of knowledge with others. The discussion contents can be converted to text by exploiting speech recognition technologies. Similarly, if one is reading a printed book, the contents of the book can be recorded by a camera like Google Glass, then converted to text by utilizing Optical Character Recognition (OCR) technology. If the book is an electronic one, no conversion is needed, text can be extracted directly.
\subsection{Analyzing the text content with topic models}
Having the extracted or converted text, with topic models, the main ideas of the text can be obtained in a quantitative manner \cite{blei2012probabilistic}. With probabilistic topic models, the main ideas of a piece of text can be computed as a distribution over a series of topics. Each topic is expressed as a word distribution over a vocabulary set. 
With the calculated topic distribution and word distribution, further analyzing of knowledge model is available.
\subsection{Analyzing the learning history with knowledge model}
Knowledge model can quantitatively evaluate an individual's knowledge based on his learning history.
\subsubsection{Organization of human knowledge}
In knowledge model, all the knowledge pieces are organized in a tree structure. Every node of the knowledge tree can be referenced by a name. A branch node represents a discipline or sub-discipline of knowledge, such as math, computer science, and information retrieval etc. A leaf node represents a concrete piece of knowledge, which is explicitly defined and has been widely accepted, such as Bayes' theorem, Mass-energy equivalence, Expectation-maximization algorithm etc. A leaf node of the knowledge tree is called a knowledge point. A branch node of it is called a knowledge branch. The knowledge tree can be constructed and maintained empirically by a group of experts of each discipline. Fig. 1 is an example of the knowledge tree based on a classification of Wikipedia\footnote{It can be found at \url{https://en.wikipedia.org/wiki/Branches_of_science}}. To keep it simple, other nodes of the tree are omitted. 

\begin{figure}[htbp]
	\centering
	\includegraphics[height=2.2in, width=4in]{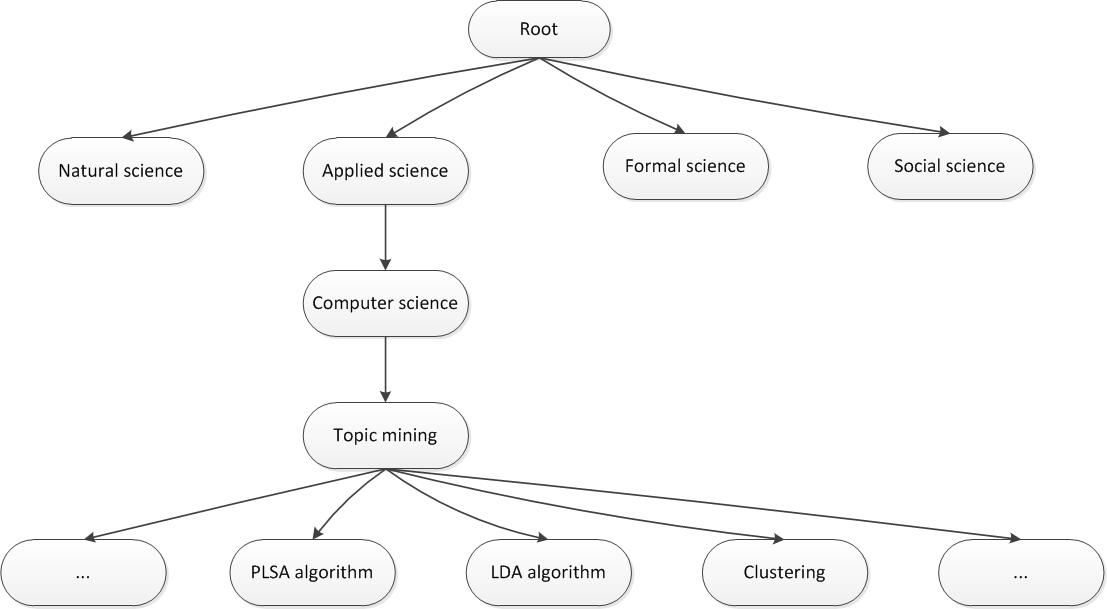}
	\caption{An example of the knowledge tree.}
\end{figure}

\subsubsection{Learning sessions}
An individual's learning activities can be separated into a series of learning sessions based on some specific standards, such as intervals between activities or topics of activities. Details of how to discriminate learning sessions will be discussed in section 3.
Table 1 illustrates some examples of learning sessions.

\subsubsection{An individual's learning history}
Each individual has a knowledge tree which records his learning history about each knowledge node. Each node of the tree has a data structure which records the individual's every learning experience about the corresponding knowledge point or knowledge branch. Each recorded learning experience has the following 4 attributes:
\begin{itemize}
\item Learning sequence ID\\
Recording the sequence ID of the learning experience.
\item Stop time\\
Recording when the learning session stopped.
\item Duration\\
Recording the duration time of a learning session.
\item Proportion\\
Recording the knowledge point's share of the learning contents. The calculation of the proportion is based on results of topic model analysis, details of calculation will be discussed in section 3.
\end{itemize}
Table 2 is an example of learning history, it is a snippet of a subject's learning history of the knowledge point ``$Bayes'~ rule$''.

\begin{table*}[htbp]
	\footnotesize
	\centering
	\caption{Some examples of learning sessions}
	\begin{tabular}{|c|c|c|c|} \hline
		Date& Activities & Duration(S)  & Captured text contents \\\hline
		... &&&  \\\hline
		\multirow{2}{20mm}{2016-03-13 9:30:00}&\multirow{2}{35mm}{Started reading a document}& \multirow{4}{*}{3610 }  &	\multirow{4}{70mm}{... Probabilistic models, such as hidden Markov
			models or Bayesian networks, are commonly ...} \\	
		&&&  \\ \cline{1-2}
		\multirow{2}{20mm}{2016-03-13 10:30:10}&\multirow{2}{35mm}{Stopped reading the document}&&  \\
		&&&  \\\hline
		
		\multirow{2}{20mm}{2016-03-13 13:30:20}&\multirow{2}{35mm}{Started attending a class}& \multirow{4}{*}{2710 }  &	\multirow{4}{70mm}{... how does the expectation maximization algorithm work ...} \\	
		&&&  \\ \cline{1-2}
		\multirow{2}{20mm}{2016-03-13 14:15:30}&\multirow{2}{35mm}{Stopped attending the class}&&  \\
		&&&  \\\hline
		
		\multirow{2}{20mm}{2016-03-13 15:10:10}&\multirow{2}{35mm}{Started a discussion}& \multirow{4}{*}{930 }  &	\multirow{4}{70mm}{... I think your understanding of Bayes' theorem is wrong ...} \\	
		&&&  \\ \cline{1-2}
		\multirow{2}{20mm}{2016-03-13 15:25:40}&\multirow{2}{35mm}{Stopped the discussion}&&  \\
		&&&  \\\hline
		
		... &&&  \\\hline
	\end{tabular}	
\end{table*}

\begin{table*}
	\centering
	\caption{A subject's learning history of the knowledge point ``$Bayes'~ rule$''}
	\begin{tabular}{|c|c|c|c|}  \hline
	Learning sequence ID&	Learning stop time&	Duration(S)&	Proportion\\ \hline
	1&	2/27/2016 18:41&	1171&	1.22\%\\ \hline
	2&	2/27/2016 18:47&	220&	2.12\%\\ \hline
	3&	2/29/2016 16:08&	2523&	1.17\%\\ \hline
	4&	2/29/2016 16:55&	330&	0.66\%\\ \hline
	5&  3/3/2016 16:21& 	1710& 	1.17\%	\\
		\hline\end{tabular}
\end{table*}

\subsubsection{Calculation of an individual's familiarity measure about a knowledge point}
With an individual's learning history of a knowledge point, it is possible to measure the individual's familiarity of the knowledge point. There is no unanimous agreement of how previous learning experiences affect an individual's current understanding of a knowledge point exactly. Therefore, there are many choices of calculating the familiarity measure. Details of calculation will be discussed in section 3. Figure 2 illustrates a flowchart of using topic model and knowledge model to estimate an individual's knowledge quantitatively. Each hexagon of the diagram indicates a step of processing, the following rectangle indicates the results of the processing. A preliminary system of evaluating an individual's knowledge is implemented in section 3. 
\begin{figure}[htbp]
	\centering
	\includegraphics[height=3.6in, width=2.1in]{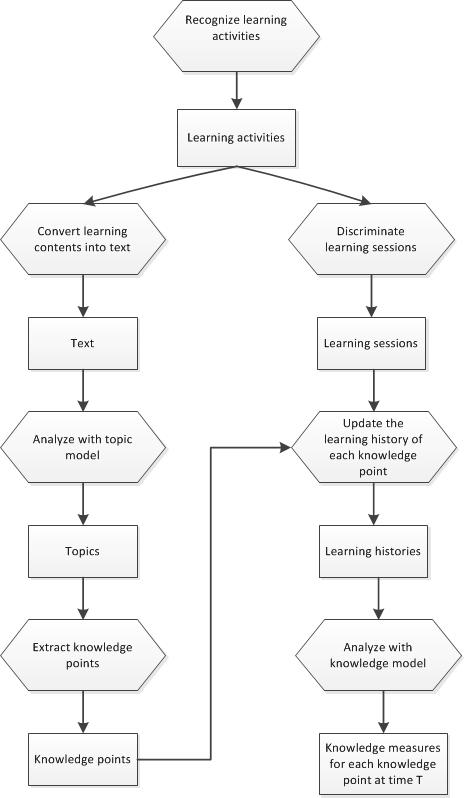}
	\caption{A flowchart to estimate an individual's knowledge quantitatively.}
\end{figure}

\section{Related works}
Recording an individual's learning history is vital for knowledge model. Bush envisioned the $ memex $ system in which individuals could compress and store all of their personally experienced information, such as books, records, and communications\cite{bush1945atlantic}.  Inspired by $ memex $, Gemmell et al. developed a project named `$ MyLifeBits $' to store all of a person's digital media, including documents, images, sounds, and videos\cite{gemmell2002mylifebits}. Knowledge model shares a similar idea with $ memex $ and `$ MyLifeBits $' of recording an individual's digital history, but with a different intention. $ Memex $ and `$ MyLifeBits $' are mainly for re-finding or review of personal data, knowledge model is for quantitatively evaluating a knowledge worker's knowledge.
\paragraph*{Probabilistic topic model.} Probabilistic topic model is used to analyze the topics of a collection of text documents. Each topic is represented as a multinomial distribution of words over a vocabulary set. Each document is represented as a distribution over the topics \cite{steyvers2007probabilistic, blei2012probabilistic}. Probabilistic latent semantic analysis (PLSA) \cite{hofmann1999probabilistic} and Latent Dirichlet Allocation (LDA) \cite{blei2003latent } are two representative probabilistic topic models. PLSA models each word of a document as a sample from a mixture model. It has a limitation that parameterization of the model is susceptible to over-fitting. In addition, it cannot provide a straightforward way to make inferences about new documents\cite{lu2011investigating}. LDA is an unsupervised algorithm that models each document as a mixture of topics. It addresses some of PLSA's limitations by adding a Dirichlet prior on the per-document topic distribution.
\paragraph*{Forgetting curve.} Human memory declines along time. In 1885, Hermann Ebbinghaus hypothesized the exponential nature of forgetting \cite{ebbinghaus1913memory}. Ebbinghaus found Equation 1 can be used to describe the proportion of memory retention after a period of time, $ t $ is the time in minutes counting from one minute before the end of the learning, $ k $ and $ c $ are two constants which equal 1.84 and 1.25 separately\footnote{It can be found at \url{http://psychclassics.yorku.ca/Ebbinghaus/memory7.htm}}.
\begin{equation}
b = k/((\log t)^c + k)
\end{equation}
Figure 3 shows the percentage of memory retention in time calculated by Equation 1.
\begin{figure}[htbp]
	\centering
	\includegraphics[height=2.4in, width=2.4in]{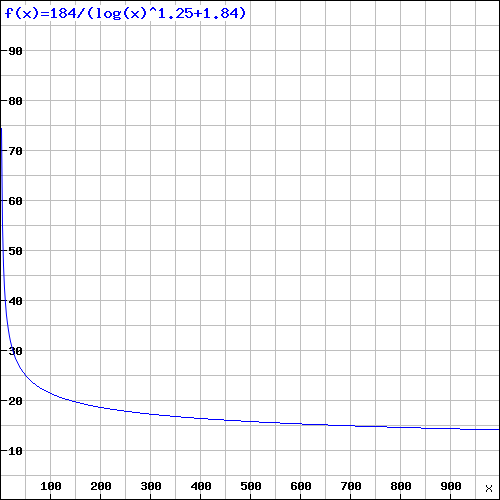}
	\caption{The percentage of memory retention in time calculated by Equation 1.}
\end{figure}
Averell and Heathcote proposed other forms of forgetting curves. There is no unanimous agreement of how human memory declines. Psychologists have debated the form of the forgetting curve for a century \cite{averell2011form}.

\section{A preliminary knowledge evaluating system}
A preliminary knowledge evaluating system is developed to test the feasibility of knowledge model. 
Because of the complexity of human learning activities and the workload of programming, it is impractical to handle all the learning situations once and for all. Therefore, it only handles the situation that a user is reading Portable Document Format (PDF) documents. Other document formats and learning methods like listening and discussing will be considered in further research.

A plug-in for the Adobe Acrobat Reader application is developed. With the plug-in, the system can detect an individual's PDF reading activities, then divides them into a sequence of learning sessions. Meanwhile, it extracts the text contents of each learning session, then uses topic model to analyze the topics of the text contents, and then selects the topics which are knowledge points, finally, it updates the individual's learning histories of related knowledge points. With the learning histories, the individual's familiarity measure of each knowledge point at time $ t $ can be calculated with knowledge model.

\subsection{An Algorithm to Discriminate Learning Sessions}
Discriminating learning sessions is critical to knowledge model because it is essential to know how many times and how long for each time the individual has learned a knowledge point. Further analyses are based on these results. Algorithm 1 is devised to discriminate learning sessions when the user is reading. The algorithm periodically checks what the individual is doing. Either of the following three conditions indicates a learning session has started.
\begin{enumerate}
	\item The individual opens a document;
	\item The foreground window has switched to an opened document from another application (APP), such as a game APP;
	\item After the computer being idled for a period of time, there are some mouse or keyboard inputs detected, which indicates the individual has come back from other things. Meanwhile, the foreground window is an opened document.
\end{enumerate}

If either of the following three conditions is satisfied, a learning session is assumed to be terminated.
\begin{enumerate}
	\item The individual closes a document;
	\item The foreground window has switched to another APP from a document;
	\item The foreground window is a document, but the computer has idled for a certain period of time without any mouse or keyboard inputs detected, the individual is assumed to have left to do other things.
\end{enumerate}

The algorithm periodically checks if any of the conditions listed above is satisfied, if so, it records a learning session has started or stopped. When a document is opened or closed, the PDF Reader APP will send the plug-in a message, so there is no need to check these two actions. The duration of a learning session equals the interval between its start and stop time. Page numbers are recorded for the purpose of extracting learning content, which will be analyzed with topic model.

\begin{algorithm}
	\caption{An algorithm to discriminate one's learning sessions when reading}
	\label{alg1}
	\begin{algorithmic}[1]			
		\WHILE{The PDF Reader APP is running}		
		\IF{The foreground window has switched to another APP from a document \textbf{OR}\\
			the computer has idled for a certain period of time when showing a document}					
		\STATE record that the document's learning session has stopped;
		\ELSE
		\IF{The foreground window has switched to an opened document from another APP \textbf{OR}\\
			the individual has come back to continue reading a document}			
		\STATE record that a learning session about the current document has started;
		\ENDIF
		\ELSE
		\IF{There is no APP and document switch}			
		\STATE check and record if there is a Page switch;
		\ENDIF			
		\ENDIF
		\STATE keep silent for T seconds;
		\ENDWHILE		
	\end{algorithmic}	
\end{algorithm}

Figure 4 shows some examples of discriminated learning sessions. Attribute ``$ did $'' means document ID, which indexes the documents uniquely. Attribute ``$ actiontype $'' indicates the type of an action. ``$ Doc~ Act $'' means a document has been activated. ``$ Page~ Act $'' is defined similarly. ``$ Doc~ DeAct $'' means a document has been deactivated. That is to say, a learning session has stopped.  Attribute ``$ page $'' indicates a page number. Attribute ``$ duration $'' records how long a page has been activated in seconds. If two learning session's interval is less than a certain threshold, such as 30 minutes, and their learning material is the same, for example, the same document, they are merged into one session. Therefore, ``$ Session 2 $'' and ``$ Session 3 $'' are merged into one session.

\begin{figure}[htbp]
	\centering
	\includegraphics[height=2.45in, width=2.8in]{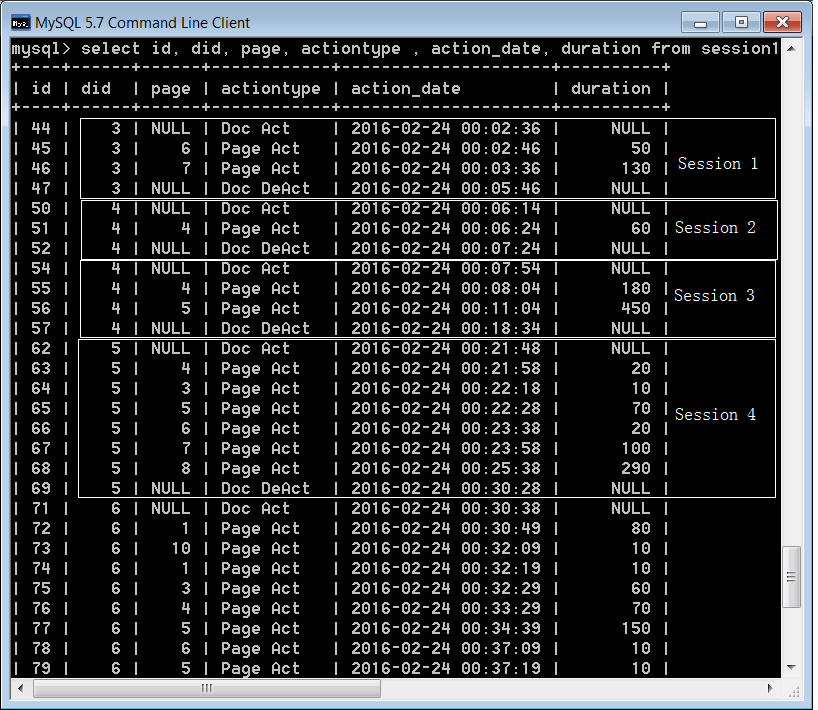}
	\caption{Some examples of discriminated learning sessions.}
\end{figure}

\subsection{Analyzing the learning contents with topic model}
When discriminating learning sessions, text learning contents are also extracted. Because Algorithm 1 can record the accurate set of pages the individual has read during a learning session, only the related pages' text contents are extracted. This strategy brings in less errors than extracting all the text contents of the whole document. Because a document may contain many pages, usually some of them are not read during a learning session, it is unreasonable to count them in. The inputs of a probabilistic topic model are a collection of $ N $ documents, a vocabulary set $ V $, and the number of topics $ k $. The outputs of a probabilistic topic model are the followings:

\begin{itemize}
	\item k topics, each is word distribution : $ \{\theta_{1},...,\theta_{k}\} $;
	\item Coverage of topics in each document $ d_{i} $: $ \{\pi_{i1},...,\pi_{ik}\} $;\\
	 $ \pi_{ij} $ is the probability of document $ d_{i}$ covering topic $ \theta_{j} $.	
\end{itemize}

In the implementation, N is set 1 because there is only one document during a learning session, k is set 2 currently. The LDA analysis of learning contents is based on the implementation of MeTA, which is an open source text analysis toolkit \footnote{The package is available at \url{https://meta-toolkit.org}}. Before topic model analysis, the text learning contents are scanned to find out the word group which is a multi-word knowledge point, such as ``inverse document frequency'' (IDF). The word group is then merged into one word like inverse-document-frequency. After the merging of multi-word knowledge points, the text contents are analyzed with the unigram method of LDA. 

\subsection{Computation of a knowledge point's share of the learning contents}
Topic model can calculate each topic's contribution to the learning contents and each term's share of a topic. Each knowledge point can be allocated with a share based on its topic share. The share is an estimation of how much the learning contents concern the knowledge point.
Only the top $ m $ terms of each topic are considered. Each related topic term's share is calculated with Equation 2. $ \varphi_{ij} $ is the share of term $ i $ of topic $ j $, $ \pi_{j} $ is topic $j $'s share of the learning contents, $ p(t_{i}|\theta_{j}) $ is term $ i $'s share of topic $ j $. A knowledge point's share equals its topic term share.

\begin{equation}
\varphi_{ij} = \frac{\pi_{j}p(t_{i}|\theta_{j})}{\sum_{j=1}^k\sum_{i=1}^m\pi_{j}p(t_{i}|\theta_{j})}
\end{equation}

\subsection{Computation of the familiarity measure of a knowledge point at a particular time}
With the recognized learning sessions and the results of topic model analysis, an individual's learning history of a knowledge point can be generated. Table 2 shows an example of an individual's learning history of a knowledge point.
With the learning history, there are many choices to calculate the individual's familiarity measure of a knowledge point. The simplest method is just considering the cumulative learning time of each knowledge point, multiplied by its corresponding share in each learning session. However, human brain works in a very complicated manner when learning. A lot of factors affect how effective an individual can learn a knowledge point. For example, human memory declines. There is much difference between learning a knowledge point yesterday and three years ago. Moreover, subsequent learning of a knowledge point will be associated with what have been learned previously. A simplified method of calculating familiarity measures is used in this preliminary implementation. The computation is based on the following hypotheses:
\begin{itemize}
\item Each learning experience of a knowledge point is independent from other learning experiences of it;
\item The effect of each learning experience declines in time according to Ebbinghaus' forgetting curve of Equation 1;
\item The familiarity measure of a knowledge point is the additive effects of all the learning experiences of it.
\end{itemize}
Equation 3 is used to calculate an individual's familiarity measure of knowledge point $ k_{i} $ at time $ t $. The input is a sequence of $ n $ learning sessions. $ d_{j} $ is session $ j $'s duration in seconds;~  $ \xi_{ij} $ is knowledge point $ k_{i} $'s share in session $ j $, it is calculated with Equation 2;~ $ b_{j} $ is the proportion of memory retention of learning session $ j $ at time $ t $, it is calculated with Equation 1.
\begin{equation}
F_{k_{i}} = \sum_{j=1}^nd_{j}*\xi_{ij}*b_{j}
\end{equation}

A relative familiarity measure can be calculated by dividing the familiarity measures with the mean value of them.
\subsection{Results}
A subject's 13 days (from 2/23/2016 to 3/6/2016) of PDF documents reading histories are recorded and analyzed. During the period of time, the subject has read 38 documents for 417 times.
For the simplicity of calculation, pages on which the subject has spent less than 30 seconds are ignored; learning sessions which are less than 150 seconds are also ignored. After the filtering, there are a total of 43 learning sessions recognized, 69 knowledge points were captured. Table 3 illustrates the subject's statistics and familiarity measures of 5 randomly selected knowledge points, the calculation time is 2016-03-29 19:24:00. The values of familiarity measures change over time, because human memory declines over time.
\begin{table*}
	\centering
	\caption{A subject's statistics and familiarity measures of 5 randomly selected knowledge points}
	\begin{tabular}{|l|c|c|l|c|} \hline
		
		\multirow{2}{*}{Knowledge point name}&	Learning&	Cumulative&	\multirow{2}{*}{Latest learning date}&	Familiarity \\
		&frequency& learning time(S)& &	measure \\ \hline
		Bayes' rule&	5&	5954&	3/3/2016 16:21&	15.14 \\ \hline
		Conditional entropy&	3&	6294&	2/24/2016 16:13&	25.75\\ \hline
		Posterior distribution&	5&	4715&	3/5/2016 17:44&	35.05\\ \hline
		Lagrange multiplier&	1&	751&	2/27/2016 19:52&	3.97\\ \hline
		Expectation-maximization&\multirow{2}{*}{12}	&\multirow{2}{*}{11448}	&\multirow{2}{*}{3/3/2016 16:21}	&	\multirow{2}{*}{122.54}\\ 
		algorithm& &	 &	 & \\
		
		\hline\end{tabular}
\end{table*}

\section{Potential applications of knowledge model}
With a quantitative evaluation of an individual's knowledge, many decisions which were made empirically can now be considered based on a numerical analysis. The following are some examples:
\subsection{Searching common topics}
As mentioned in section 1, knowledge model can be used to discover common topics efficiently for people with different education or cultural background. A discipline or sub-discipline they both are interested in can be selected first, then find out the set of knowledge points with which they both are familiar based on the familiarity measures. These knowledge points can serve as the common topics of their conversation. This application can be extended to discover common topic terms which are not defined as knowledge points, such as a movie star's name.
\subsection{Selecting a lecture}
It is common for a knowledge worker to take part in all kinds of academic lectures. It is frustrated and wasting of time that a lecture is too recondite to understand. To help a potential audience predict how much he can understand the contents of the lecture, the lecturer can list a set of knowledge points which are important to understand it on the poster, then the audience can check his familiarity measures of those knowledge points. A score of how much he can understand it can be calculated based on the familiarity measures.
\subsection{Evaluating a scientist's research concentrations in a period of time}
Because an individual's learning histories of all knowledge points have been recorded. It is convenient to extract fragments of the learning histories of a period of time, for example, the last three years, then use knowledge model to calculate the familiarity measures of that period. The set of knowledge points which have larger familiarity measures are the scientist's research concentrations.
\subsection{Selecting appropriate referees for a research paper}
When a research paper is submitted for reviewing, choosing the optimal referees from a candidate set is a difficult problem. At present it is usually decided empirically. With knowledge model, an objective numerical analysis is possible. For example, each candidate referee's research concentrations can be calculated, the submitted paper's knowledge points and their corresponding shares can also be calculated, by matching these values, the optimal referee list can be obtained.
\subsection{Evaluating a knowledge worker's expertise on a discipline or sub-discipline}
With an individual's familiarity measures of all the knowledge points, it is not hard to evaluate his expertise on a discipline or sub-discipline. The knowledge points are organized in a tree structure, each subtree represent a discipline or sub-discipline of knowledge. The evaluation can be made based on how many knowledge points the individual has mastered and the average familiarity measure of the subtree. 

\section{Conclusion}
In this paper, a method named knowledge model which can quantitatively evaluate a knowledge worker's knowledge is proposed. The main idea is to record an individual's learning histories of each piece of knowledge, and then use the learning history as an input to calculate the individual's familiarity measure of each knowledge point. A preliminary knowledge evaluating system is developed, it analyzes an individual's PDF documents reading activities, then uses topic model and knowledge model to calculate the individual's familiarity measures of captured knowledge points. An algorithm of discriminating learning sessions is devised. In addition, a method of calculating the individual's familiarity measure of a knowledge point based on its learning history is proposed.

\section{Discussion}
Evaluating a person's possession of knowledge is very complicated. This part discusses related issues about individual knowledge evaluation.
\subsection{Cognitive assumption of knowledge model}
Knowledge model focuses on evaluating a person's gaining of conceptual knowledge. Because it is difficult for a machine to observe learning of procedural knowledge. However, since learning of procedural knowledge is usually companioned by learning of conceptual knowledge, it has some ability for assessing the gaining of procedural knowledge. Knowledge model divides a person's learning activities into two categories: Observable Learning Activities (OLA) and Unobservable Learning Activities (ULA). For OLA, the learning start and stop time, and the text contents of it are observable, such as reading, listening, discussing, writing, and speaking. For ULA, the learning start and stop time, and the text contents of it are unobservable, such as rumination and meditation. Therefore, we cannot analyze ULA.

If the interval between the evaluation time and the last learning time of a knowledge point is large, the third variable of Equation 3 is approximating to a constant. Because the speed of memory decay is attenuating. If we set the third variable of Equation 3 to a constant and summarize all the knowledge points' familiarity measures, we obtain a value that is proportional to a person's total time spent on learning; that is to say, knowledge model assumes the quantity of a person's knowledge is proportional to his/her total time spent on learning knowledge. This quantity is then distributed to different knowledge points. A set of related knowledge points forms a domain. Since ULA are unobservable, the quantity and distribution of ULA are also unobservable. Knowledge model does not make assumption about the total quantity of ULA. However, it assumes the distribution of ULA is equivalent to OLA (at least in domain level). To put simply, knowledge model does not assume how much time a person has spent on rumination; it assumes the contents of ruminations are related to, and are proportional to (at least in domain level) what the person has experienced. E.g., a farmer who has nothing to do with quantum physics cannot regularly ruminate about the topics in quantum physics, contrarily, a quantum physicist will do; a person cannot ruminate a concept he has never seen or heard, unless he is the creator of the concept.

\subsection{Normalization among knowledge points}
The calculation of familiarity measure mainly considers the individual's time devotion to a knowledge point and its share of each learning content. However, the complexity levels of knowledge points are usually different. For example, spending 20 minutes is sufficient for a normal knowledge worker to understand and remember the Pythagorean Theorem, but it is usually not enough to understand a complicated algorithm like LDA. Therefore, the familiarity measures should be normalized among knowledge points. Each knowledge point can be allocated with a complexity level. The familiarity measure can be multiplied by a factor, which is a function of the knowledge point's complexity level. The complexity level of a knowledge point can be decided empirically by a group of experts when constructing the knowledge tree. Another method for calculating complexity level is by examining its Understanding Graph \cite{abs-1711-06553}. Another method for normalizing familiarity measures among knowledge points is by analyzing its Understanding Tree \cite{1612.07714}. If we want to evaluate a person's possession of knowledge about a domain, the average familiarity measure of that domain can be used.

\subsection{Normalization among knowledge workers}
If knowledge model is used for self-evaluating, like most applications mentioned in section 4, it is not essential to normalize familiarity measures among knowledge workers. A standardized value of subtracting the mean value and then divided by the standard deviation is sufficient. If knowledge model is used for making decisions for a competition, such as using knowledge model analysis as a substitution of an examination (test), normalization of familiarity measures among knowledge workers is essential. The normalization can be made by multiplying the familiarity measures by a factor that is determined by the subject's characteristics, such as Intelligence Quotient (IQ). 

According to \cite{hunt2010human}, human IQs are normally distributed with a mean value of 100, and standard deviation of 15. Approximately two-thirds of all scores lie between 85 and 115. Five percent (1/20) are above 125, and one percent (1/100) are above 135. Similarly, five percent are below 75 and one percent below 65. Therefore, in many circumstances, such as evaluating a group of undergraduates from the same university and department, we can hypothesize the people's IQs are equivalent. Even under this assumption, we cannot compare people's familiarity measures directly, because the existence of ULA; unless the gap between familiarity measures is distinct. 

The factor can be determined by test. E.g., by testing, we conclude person A's familiarity measure of 100 is equivalent to person B's familiarity measure of 80. If we define person A's factor is 1, then person B's factor is 1.25. Cheating is inevitable in a lot of competitions, it can be detected by sampling a set of knowledge points and examining them by test.

\subsection{Using topic as knowledge unit}
Knowledge model uses knowledge point (concept) as a unit of knowledge. A knowledge point is defined as a piece of knowledge that is explicitly defined and has been widely accepted, it is embodied by a concept. This definition makes an attempt to exclude concepts that are unsubstantiated, such as the ones presented in a newly published paper. The name conveys a thinking that a ``knowledge point" is just a ``point" in the tremendous tree or network of knowledge. It is trivial or less meaningful to examine one isolated knowledge point; it is more meaningful to analyze a group of related knowledge points, such as all the knowledge points of a domain, or knowledge points that are organized by Understanding Graph \cite{abs-1711-06553} or by a topic of topic model.

A topic of topic model can also be considered as a unit of knowledge \cite{wallach2006topic,huang2016framework}. It has the advantage that a topic is naturally a set of related concepts and the topics can be calculated automatically by machines. The disadvantage is that the topics are unfixed; when the corpus changes, the topics also change. This characteristic makes it  unsatisfactory as a unit of knowledge. If we choose a topic of topic model as a unit of knowledge, Equation 4 can be used to calculate a person's familiarity measure to topic $ T_{i} $ at time $ t $. The input is a sequence of $ m $ learning sessions that are related to topic $ T_{i} $. $ d_{j} $ is session $ j $'s duration;~  $ \rho_{ij} $ is topic $ T_{i} $'s share in session $ j $, it is calculated by topic model;~ $ b_{j} $ is the proportion of memory retention of learning session $ j $ at time $ t $, it is calculated with Equation 1. Each document in the corpus matches the text contents of a learning experience.

\begin{equation}
F_{T_{i}} = \sum_{j=1}^md_{j}*\rho_{ij}*b_{j}
\end{equation}

\subsection{Constructing concepts pool}
A concepts pool is defined as a set of concepts that are select based on some standards. The following lists several types of concepts pool:
\begin{itemize}
	\item Type 1 concepts pool is corpus based, the concepts are selected by analyzing a corpus.
	\begin{itemize}
		\item Type 1A concepts pool is constructed by checking the Term Frequency (TF) of a concept in a corpus. If the TF is larger than a threshold, the concept is selected.
		\item Type 1B concepts pool is constructed by checking the Inverse Document Frequency (IDF) of concepts.
	\end{itemize}
	\item Type 2 concepts pool is familiarity measure based, the concepts are selected by checking a person's or a group of people's familiarity measures at sometime.
	\begin{itemize}
		\item Type 2A concepts pool is constructed by checking a person's familiarity measure of a concept at time $ t $. If the familiarity measure is larger than a threshold, the concept is selected.
		\item Type 2B concepts pool is constructed by checking a group of people's familiarity measures at sometime.
	\end{itemize}
	\item Type 3 concepts pool is based on the structure of knowledge. Such as selecting concepts from a concept's n-level neighborhood in an Understanding Map \cite{abs-1711-06553}, or a concept map \cite{mcclure1999concept}.
\end{itemize}
By comparing a Type 1A concepts pool with a Type 2A, a person's expertise in a domain can be obtained. By examining a Type 2B concepts pool, some cultural elements that belong to a society or nation of people can be obtained.

\subsection{Using logistic regression for estimating understanding}
In \cite{1612.07714}, the average familiarity measure in an Understanding Tree (except for the root) is used to estimate a person's understanding degree to the root concept. The root is differentiated to make sure the subject has substantial learning experiences about it. Equation 2 of \cite{1612.07714} assumes the effect of each descendant to the understanding of the root is equal. However, different descendants may play different roles for understanding the root. Logistic regression can be used to discriminate effects of different descendants. Equation 5 is the logistic regression equation. $ F_{k_{j}}(t) $ is a concept's familiarity measure on the understanding tree at time $ t $, $ \alpha_{j} $ is its coefficient. The parameters can be determined by test. Equation 6 calculates the subject's probability of understanding the root at time $ t $. Besides selecting concepts from an Understanding Tree for evaluation, an alternative method is selecting from the root's n-level neighborhood in an Understanding Map \cite{abs-1711-06553}.

\begin{equation}
\theta (t) = \alpha_{0} + \sum_{j=1}^m \alpha_{j} * F_{k_{j}}(t)
\end{equation}

\begin{equation}
P_{r}(t) = \frac{1}{1 + e^{-\theta (t)}}
\end{equation}

\subsection{Evaluation of knowledge model}
Evaluation of the effectiveness of knowledge model is a huge project. 
\begin{itemize}
\item Firstly, a knowledge worker's almost all the learning activities should be recorded and analyzed, such as reading of all kinds of documents and web pages, attending of lectures, oral discussions etc. Each knowledge point's complete learning history is obtained, its relative familiarity measure is also computed; 
\item Secondly, select a sample of knowledge points and group them according to their relative familiarity measures; 
\item Thirdly, let the individual take an examination which tests his understanding of the sample knowledge points; 
\item Fourthly, compare the results of the examination with the relative familiarity measures calculated by knowledge model; 
\item Finally, repeat the above procedures on other knowledge workers to reduce the randomness of the results. 
\end{itemize}
A detailed evaluation will be considered in further research. 
\subsection{Limitations of using Ebbinghaus' forgetting curve}
Ebbinghaus' forgetting curve formula is used in the computation of familiarity measures. It depicts the decline of memory retention in time. Many research results have testified the soundness of the formula \cite{murre2015replication, rubin1996one}. However, other factors may affect the speed of memory decay as well. Such as how the information is presented and the physiological state of the individual. There are no unanimously accepted formulas of how these factors affect the speed of memory decay. In addition, it is difficult to obtain accurate values for these factors. 

The calculation of the familiarity measures is based on the individual's learning histories of a long range of time, usually several years or decades of years. In my opinion, when observing from a long time range, it can be hypothesized that the average presentation qualities and average physiological states among knowledge points are equivalent , so these factors can be ignored. If other forms of forgetting curve formulas are proved to be better than Ebbinghaus', it can be used as a substitution when calculating familiarity measures.
\subsection{Privacy issues}
Recording learning histories of each knowledge point will inevitably violate an individual's privacy. To protect the privacy, the learning histories can be password protected or even encrypted. They are stored in the individual's personal storage, and should not be revealed to other people. The only information the outside world can see is the individual's familiarity measures of the knowledge points. The knowledge points which may involve the individual's privacy are separated from other knowledge points, every output of their familiarity measures should be authorized by the owner.

\footnotesize
\bibliography{kmodel}

\bibliographystyle{abbrv}

\end{document}